\documentclass[preprint,aps,prb,showpacs,showkeys]{revtex4}

\usepackage{graphicx}
\begin{document}
\draft
\title{Localization Properties of Quantized Magnetostatic Modes in Nanocubes}
\author{H. Puszkarski\cite{correspond}, M. Krawczyk}
\address{Surface Physics Division, Faculty of Physics, Adam
Mickiewicz University, \\ ul. Umultowska 85, Pozna\'{n}, 61-614
Poland.}
\author{J.-C. S. L\'{e}vy}
\address{Laboratoire de Physique Th\'{e}orique de
la Mati\`{e}re Condens\'{e}e, case 7020,\\ Universit\'{e} Paris 7,
2 place Jussieu, 75251 Paris C\'{e}dex 05 France.}

\date{\today}

\begin{abstract}
We investigate the dynamical properties of a system of interacting
magnetic dipoles disposed in sites of an \textit{sc} lattice and
forming a cubic-shaped sample of size determined by the cube edge
length $(N-1)a$ ($a$ being the lattice constant, $N$ representing
the number of dipolar planes). The dipolar field resulting from
the dipole-dipole interactions is calculated numerically in points
of the axis connecting opposite cube face centers (\textit{central
axis}) by collecting individual contributions to this field coming
from each of the $N$ atomic planes perpendicular to the central
axis. The applied magnetic field is assumed to be oriented along
the central axis, magnetizing uniformly the whole sample, all the
dipoles being aligned parallelly in the direction of the applied
field. The frequency spectrum of magnetostatic waves propagating
in the direction of the applied field is found numerically by
solving the Landau-Lifshitz equation of motion including the local
(\textit{nonhomogeneous}) dipolar field component; the mode
amplitude spatial distributions (\textit{mode profiles}) are
depicted as well. It is found that only the two energetically
highest modes have \textit{bulk-extended} character. All the
remaining modes are of \textit{localized} nature; more precisely,
the modes forming the lower part of the spectrum are localized in
the \textit{subsurface region}, while the upper-spectrum modes are
localized around the sample center. We show that the mode
localization regions narrow down as the cube size, $N$, increases
(we investigated the range of $N$=21 to $N$=101), and in
sufficiently large cubes one obtains practically only
\textit{center-localized} and \textit{surface-localized}
magnetostatic modes.
\end{abstract}

\keywords{magnetostatic modes; magnetic nanograins; mode
localization; demagnetizing field}

\pacs{75.30.Ds, 75.40.Gb, 75.75.+a}

\maketitle

\section{Introduction}
The development of nanotechnology in recent years allowed to
design nanometric ferromagnetic materials of any required shape
with a large size variation range \cite{1}. Small magnets and
particles have raised an increasing interest due to their
potential application in magnetic random access memory (MRAM)
elements and ultrahigh-density storage. This interest also
involves the sample ability to rapidly change the magnetization
state in order to optimize the individual recording  time.
Particularly, magnetostatic modes are of practical importance in
this matter (especially in magnetization reversal processes), by
reason of their low frequency range.

The small magnetic elements used in signal processing devices are
of nanoscale size and mostly non-ellipsoidal shape. In such small
magnetic particles the internal field non{\-}uniformity due to the
particle shape has to be taken into account. Of course, the
dominant mechanism in the creation of such a nonhomogeneity are
long-range dipolar interactions. The dipolar size and shape
effects present in the magnetostatic wave spectra of small
nonellipsoidal particles result in additional power losses
\cite{2} very important in device applications.

The nonuniform demagnetizing field in a normally magnetized disk
was studied long ago by Yukawa and Abe \cite{3}, with the
assumption of uniform saturation magnetization throughout the
disk. Analyzing the ferromagnetic resonance (FMR) spectrum of the
magnetostatic waves in a normally magnetized YIG disk, the authors
found that these waves were excited at the disk edge and
propagated radially towards the disk center with increasing wave
numbers.

The dynamical properties of small magnetic elements have been
actively investigated recently in studies using different methods
and approaches. Resonance methods combined with surface tunneling
microscopy techniques enabled several teams \cite{4,5,6} to
observe magnetostatic and exchange modes on a very local scale;
time-resolved and space-resolved MOKE experiments \cite{7} were
used to study the local magnetization dynamics as well. These
different methods give good indications of the localized
magnetostatic mode existence, especially when the applied field is
strong enough to induce a magnetization saturation throughout the
sample.

The localized magnetostatic mode dynamics has been investigated by
several authors. Berkov {\em et al.} \cite{8} studied the spin
wave frequency spectrum and spatial distribution in thin $\mu
$m-sized magnetic film samples. The lowest-frequency modes were
found to correspond to oscillations restricted to the boundary
regions, which start to `propagate' inside a rectangular magnetic
sample as the frequency increases. Tamaru {\em et al.} \cite{9}
applied a spatially resolved FMR technique to quantized
magnetostatic mode imaging in small magnetic structures. They
identified quantized magnetostatic modes in the obtained spectra,
and from the observed spatial distribution of magnetization
response at each mode peak deduced that the number of mode nodes
decreased with increasing bias field. However, in their data
interpretation some discrepancies were found between the measured
mode frequency values and those calculated on the basis of the
Damon-Eshbach theory \cite{10}. An exact theory of spin wave mode
quantization in small magnetic structures needs to be developed.

The dipole-exchange spectrum of the discrete spin wave modes
resulting from the bias magnetic field inhomogeneity inside a
nonellipsoidal rectangular magnetic sample has been studied
recently by Guslienko {\em et al.} \cite{11}. The authors found
that the strong inhomogeneity of the internal bias magnetic field
along the magnetization direction results in spin wave mode
localization at either the edges (exchange localization) or the
center (dipolar localization) of the sample. We will show in this
paper that in a cubic-shaped sample with an internal field
nonhomogeneity resulting solely from pure dipolar interactions
these two types of magnetostatic mode localization are also
present (it means magnetostatic modes are found to be localized
not only at the cube center, but also on the cube surfaces).

Our theoretical approach is based on a modified Draaisma-de Jonge
treatment \cite{12} used in the dipolar energy calculation. In the
original Draaisma-de Jonge treatment a ferromagnetic film is
considered as a set of discrete magnetic dipoles regularly
arranged in a crystalline lattice. The dipolar energy is
calculated by collecting contributions from each dipolar lattice
plane assumed to be parallel to the film surface. The dipoles
within each plane are divided into two sets: those within a circle
of radius $R$ and those beyond that circle; the total contribution
of the former set is calculated discretely through summing over
the dipoles, while the contribution of the latter set is evaluated
by integration (the circle radius should be large enough to assure
reliable final results). In the approach used in our study, the
discrete summation is performed over \textit{all }the dipoles
within a given dipolar plane; thus, the integration is avoided,
and no approximation is involved in our dipolar energy evaluation.

The reason to use a discrete approach in the dipolar field
calculation is our general belief \cite{13} that the continuous
models \cite{10} used in deriving magnetostatic modes become
rather questionable when applied to small samples containing a
finite number of elementary spins. The complexity of introducing
the long-range spin-spin interactions into the discrete approach
is balanced by the restriction of our study only to the case of a
uniform mode within each dipolar plane, which reduces the 3D
problem to a single dimension (this assumption is experimentally
justified by the observed weak in-plane variability of the
magnetic modes \cite{14}).

\section{A Finite System of Magnetic Dipoles in Planar
Arrangement }\label{theory}

We shall consider a system of magnetic moments $\mu_{\vec{r}}$
arranged regularly in sites $\vec{r}$ of a simple cubic crystal
lattice. The system is assumed to form a rectangular prism with
square base (see Fig. \ref{konfprost}a). Let the prism base
determine the $(x, y)$-plane of a Cartesian reference
 system, with the z-axis perpendicular to this plane. The reference point $(0, 0, 0)$ shall be
 placed in the {\em central} site of the prism bottom.

Let us calculate magnetic field $\vec{h}_{\vec{R}}$  "produced" by
all the prism dipoles in a site indicated by internal vector
$\vec{R}$. According to the classical formula (obtained using the
linear approximation \cite{landau}), field $\vec{h}_{\vec{R}}$ can
be expressed as follows (in the SI units):
\begin{equation}
\label{eq5}
\vec {h}_{\vec{R}} = \frac{1}{4 \pi} \sum\limits_{\vec{r} \neq \vec{R}}
 {\frac{3(\vec{r}-\vec{R}) \left(\vec{\mu}_{\vec{r}} \cdot (\vec{r} -\vec{R})
 \right)-\vec{\mu}_{\vec{r}} |\vec{r}-\vec{R}|^{2}}{|\vec{r}-\vec{R}|^{5}}},
\end{equation}
the above sum involving all the sites {\em except} the reference
point ({\em i.e.} the site with position vector $\vec{r} \equiv
\vec{R}$). The lattice planes parallel to the prism base shall be
numbered with index $n \in \langle 0, N-1\rangle$ (see Fig.
\ref{konfprost}b), and the sites within each plane indexed with
vector $\vec{r}_{||}=a[p\hat{i}+q\hat{j}]$, defined by integers
$p,q \in \langle -L,L\rangle $. This means that the position of
the sites in which the magnetic moments are located, indicated by
vector $\vec{r}$, shall be defined by a set of three integers,
$(p, q, n)$:
\begin{eqnarray}
\vec{r} \equiv  [\vec{r}_{||},a n] \equiv a[p,q,n], \;\;\; p,q \in
\langle -L,+L\rangle \;\;\; \mbox{and}\;\; n \in\langle
0,N-1\rangle , \label{pq}
\end{eqnarray}
$a$ denoting the lattice constant. Thus, the considered prism
contains $N(2L + 1)^{2}$ magnetic moments. Below we shall focus on
the magnetic field on the $z$-axis only, assuming its direction to
be solely allowed for magnetic wave propagation. Hence, we put
$\vec{R} \equiv a[0,0,n']$, where $n' \in \langle 0,N-1\rangle $,
and re-index the dipole field: $\vec{h}_{n'} \equiv
\vec{h}_{\vec{R}}$. By vector decomposition:
\begin{eqnarray}
\vec{r}-\vec{R}=a \left( p\hat{i}+q \hat{j} +(n-n')
\hat{k}\right),
\end{eqnarray}
equation (\ref{eq5}) becomes :
\begin{eqnarray}
\vec {h}_{\vec{R}} \equiv \vec {h}_{\vec{n'}} = \frac{1}{4 \pi}
\sum\limits_{p,q,n} {}^\prime \left[ \frac{3 a(p\hat{i} +
q\hat{j}+(n-n')\hat{k}) \left (\vec{\mu}_{\vec{r}} \cdot
a(p\hat{i}+q\hat{j}+(n-n')\hat{k}) \right)
}{(a\sqrt{p^{2}+q^{2}+(n-n')^{2}})^{5}} \right. \nonumber \\
\left. - \frac{
 \vec {\mu}_{\vec{r}} a^{2}(p^{2}+q^{2}+(n-n')^{2})}{
(a\sqrt{p^{2}+q^{2}+(n-n')^{2}})^{5}} \right] \nonumber
\end{eqnarray}
\begin{eqnarray}
=\frac{1}{4 \pi} \sum\limits_{p,q,n} {}^\prime\left[
3\frac{\hat{i}(\mu_{\vec{r}}^{x} p^{2} + \mu_{\vec{r}}^{y}pq
+\mu_{\vec{r}}^{z}p(n-n'))+\hat{j}(\mu_{\vec{r}}^{x} pq +
\mu_{\vec{r}}^{x}q^{2}+\mu_{\vec{r}}^{z}q(n-n'))}{a^{3}(p^{2}+q^{2}+
(n-n')^{2})^{\frac{5}{2}}}\right. \nonumber \\
+\left. \frac{3\hat{k}(\mu_{\vec{r}}^{x} p(n-n') +
\mu_{\vec{r}}^{y}q(n-n')+\mu_{\vec{r}}^{z}(n-n')^{2})
}{a^{3}(p^{2}+q^{2}+(n-n')^{2})^{\frac{5}{2}}}
\right. \nonumber \\
- \left. \frac{(\mu_{\vec{r}}^{x}\hat{i}+\mu_{\vec{r}}^{y}\hat{j}+
\mu_{\vec{r}}^{z}\hat{k})(p^{2}+q^{2}+(n-n')^{2})}{a^{3}(p^{2}+q^{2}+(n-n')^{2})^{\frac{5}{2}}}
\right], \label{sum2}
\end{eqnarray}
the triple sum over indices $p,q,n$ $\left( \sum\limits_{\vec{r}}
\equiv \sum\limits_{n,p,q} \right)$ replacing the single sum over
vectors $\vec{r}$; the summing range is as defined in (\ref{pq}),
and the prime ($'$) at the sum symbol means that the reference
point $[0,0,n']$ is excluded from the sum. As the site
distribution within each plane $n$ is symmetric with respect to
site $(0,0,n)$, each site $(p, q)$ has its counterpart $(-p, -q)$.
Consequently, in sum (\ref{sum2}), all the terms in which $p$ and
$q$ appear in odd powers are compensated, and (\ref{sum2})
becomes:
\begin{eqnarray}
\vec {h}_{n'} = \frac{1}{4 \pi} \sum\limits_{p,q,n} {}^\prime
\left[ 3\frac{\hat{i}\mu_{\vec{r}}^{x} p^{2} + \hat{j}\mu_{\vec{r}}^{y}q^{2}+
\hat{k}\mu_{\vec{r}}^{z}(n-n')^{2}
-\hat{i}\mu_{\vec{r}}^{x}(p^{2}+q^{2}+(n-n')^{2})}{a^{3}(p^{2}+q^{2}+(n-n')^{2})^{\frac{5}{2}}}\right. \nonumber \\
\left. -\frac{
\hat{j}\mu_{\vec{r}}^{y}(p^{2}+q^{2}+(n-n')^{2})+\hat{k}\mu_{\vec{r}}^{z}(p^{2}+
q^{2}+(n-n')^{2})}{a^{3}(p^{2}+q^{2}+(n-n')^{2})^{\frac{5}{2}}} \right]
\nonumber
\end{eqnarray}
\begin{eqnarray}
 = \frac{1}{4 \pi}
\sum\limits_{p,q,n} {}^\prime\left[
\frac{\hat{i}\mu_{\vec{r}}^{x}(2p^{2}-q^{2}-(n-n')^{2})+
\hat{j}\mu_{\vec{r}}^{y}(-p^{2}+2q^{2}-(n-n')^{2})
}{a^{3}(p^{2}+q^{2}+(n-n')^{2})^{\frac{5}{2}}}
\right. \nonumber \\
+     \left.\frac{
\hat{k}\mu_{\vec{r}}^{z}(-p^{2}-q^{2}+2(n-n')^{2})}{a^{3}(p^{2}+q^{2}+(n-n')^{2})^{\frac{5}{2}}}
\right]. \label{eq6}
\end{eqnarray}
Further calculations shall be performed using double sums over $p$
and $q$, defined as follows:
\begin{eqnarray}
I_{n,n'}=\sum_{p,q}{}^\prime\frac{\left(n-n'\right) ^{2}}{\left[
p^{2}+q^{2}+\left(n-n'\right) ^{2}\right] ^{\frac{5}{2}}}
\label{I}
\end{eqnarray}
and
\begin{eqnarray}
J_{n,n'} & = & \sum_{p,q}{}^\prime\frac{p^{2}}{\left[
p^{2}+q^{2}+\left(n-n'\right) ^{2} \right] ^{\frac{5}{2}}} \equiv
\sum_{p,q}{}^\prime\frac{q^{2}}{\left[
p^{2}+q^{2}+\left(n-n'\right) ^{2}\right] ^{\frac{5}{2}}}  \nonumber \\
& \equiv & \frac{1}{2}\sum_{p,q}{}^\prime\frac{p^{2}+q^{2}}{\left[
p^{2}+q^{2}+\left(n-n'\right) ^{2}\right] ^{\frac{5}{2}}}
;\label{J}
\end{eqnarray}
the equality of the two first sums in (\ref{J}) is a consequence
of the fact that the $z$-axis is a fourfold symmetry axis for each
lattice plane (see Fig. \ref{konfprost}). Additionally, we shall
assume that all the magnetic moments within {\em a single plane}
$n$ are identical, {\em i.e.}:
\begin{eqnarray}
\vec{\mu}_{n} \equiv \vec{\mu}_{[p,q,n]}, \mbox{\hspace{0.5cm for
any }} \;\; p\;\; \mbox{and} \;\;q. \label{mun}
\end{eqnarray}
By including relations (\ref{I}-\ref{mun}) into (\ref{eq6}),
magnetic field $\vec{h}_{n'}$
 can be expressed as follows:
\begin{eqnarray}
 \vec{h}_{n'}= \frac{1}{4 \pi}
\sum\limits_{n} \left[(J_{n,n'}-I_{n,n'})
\frac{\hat{i}\mu_{n}^{x}+ \hat{j}\mu_{n}^{y}-2
\hat{k}\mu_{n}^{z}}{a^{3}} \right]. \label{eq6b}
\end{eqnarray}
Note that by assuming $ \vec{\mu}_{n} \equiv \vec{\mu}_{\vec{r}}$
all the magnetic excitations propagating in {\em plane} $(x, y)$
are excluded from our analysis, and $z$ axis becomes the only
direction of propagation allowed.

In order to obtain a simpler expression of $\vec{h}_{n'}$ we introduce a symmetric
matrix whose elements, $D_{n,n'}$, are defined as follows:
\begin{eqnarray}
D_{n,n'} \equiv J_{n,n'} - I_{n,n'} = \sum_{p,q}{}^\prime
\frac{\frac{1}{2}\left(p^{2}+q^{2}\right) - (n-n')^{2}}{\left[
p^{2}+q^{2}+\left(n-n'\right) ^{2}\right] ^{\frac{5}{2}}}.
\label{L}
\end{eqnarray}
With this matrix, the magnetic field reads:
\begin{eqnarray}
 \vec{h}_{n'}= \frac{1}{4 \pi}
\sum\limits_{n}\left[D_{n,n'} \frac{\hat{i}\mu_{n}^{x}+
\hat{j}\mu_{n}^{y}-2 \hat{k}\mu_{n}^{z}}{a^{3}} \right]. \label{eq6c}
\end{eqnarray}
It is convenient to introduce here the notion of {\em
magnetization}, a phenomenological quantity, which in the
considered case of simple cubic lattice can be defined as follows:
\begin{eqnarray}
\vec{M}_{n}=\vec{\mu}_{n}/a^{3}.
\end{eqnarray}
Then, (\ref{eq6c})  becomes:
\begin{eqnarray}
\label{eq7a} \vec {h}_{n'} = \frac{1}{4\pi}
\sum\limits_{n}D_{n,n'} \left[\hat{i}M_{n}^{x}+
\hat{j}M_{n}^{y}-2 \hat{k}M_{n}^{z} \right].
\end{eqnarray}
It should be remembered that site $(0,0,n')$ is excluded from the
sums appearing in the  equation (\ref{L}); to avoid any ambiguity,
this shall be noted explicitly by rewriting (\ref{eq7a}) in its
expanded form:
\begin{eqnarray}
\vec{h}_{n'} = \frac{1}{4\pi}\sum\limits_{n \neq n'} \sum_{p,q}
\frac{\frac{1}{2}\left(p^{2}+q^{2}\right)-(n-n')^{2}}{\left[
p^{2}+q^{2}+\left(n-n'\right) ^{2}\right] ^{\frac{5}{2}}}
 \left[\hat{i}M_{n}^{x}+\hat{j}M_{n}^{y}-2 \hat{k}M_{n}^{z} \right] \nonumber \\
 +\frac{1}{8\pi} \sum_{p,q}{}^\prime
\frac{1}{\left[ p^{2}+q^{2}\right] ^{\frac{3}{2}}}
 \left[\hat{i}M_{n'}^{x}+\hat{j}M_{n'}^{y}-2 \hat{k}M_{n'}^{z}
 \right].
\label{eq8b}
\end{eqnarray}
The second term on the right of Eq. (\ref{eq8b}) has been
calculated for $n = n'$, and thus site $\vec{r}_{||}=(0,0)$
 should be excluded from the sum over $p$ and $q$ (as indicated by the prime at the sum symbol).
Equation (\ref{L}), defining elements $D_{n,n'}$, can be
interpreted as the definition of a
 {\em dipolar} matrix $\hat{D}$ composed of these elements; as we shall see this
 matrix shall play an important role in
deducing properties of magnetic modes. As a consequence of
(\ref{L}) matrix $\hat{D}$ is symmetric, {\em i.e.} $D_{n,n'}
\equiv D_{n',n}$. Let us introduce a new variable, defined as
follows:
\begin{eqnarray}
\delta= n - n';
\end{eqnarray}
$\delta$ measures the distance between planes $n$ and $n'$. Consequently, we can write:
\begin{eqnarray}
D_{n,n'} \equiv  D_{\delta} = D_{n',n} \equiv D_{-\delta};\\
D_{\delta} = \sum_{p,q}{}^\prime
\frac{\frac{1}{2}\left(p^{2}+q^{2}\right) - \delta^{2}}{\left[
p^{2}+q^{2}+\delta^{2}\right] ^{\frac{5}{2}}}. \label{DD}
\end{eqnarray}

Note that the considered system consists of a finite number of
planes (index $n$ taking values $n = 0,1,2,\cdots,N-1$), so the
set of values available to $\delta$ depends on the
 reference plane, $n'$, with respect to which the distance is measured. However, the following
 condition must always be satisfied:
\begin{eqnarray}
0\leq n'+\delta \leq N-1. \label{DD2}
\end{eqnarray}
A sum over {\em all}  the system planes shall be in use below;
this sum shall be denoted as:
\begin{eqnarray}
\sum_{\delta} {}^{n'},
\end{eqnarray}
superscript $n'$ indicating that the summing is performed on the planes neighbouring with $n'$,
including plane $n'$. Therefore, $\delta$ takes the following values:
\begin{eqnarray}
\delta = 0, \pm 1, \pm 2\, ...
\end{eqnarray}
its lower and upper limits being determined by condition
(\ref{DD2}).

Up to now, the direction of dipole arrangement has not had much
importance in our reasoning. Now we shall consider the case with
dipoles arranged along the $z$-axis only.

\section{ Magnetic dipoles aligned along the direction of magnetostatic mode propagation}

In this paragraph we shall consider a magnetic prism placed in a
static magnetic field, $H_{0}$, applied along the $z$-axis (Fig.
\ref{konfprost}). Field $H_{0}$ is assumed to be strong enough to
arrange {\em all} the magnetic moments along the $z$-axis. Then,
the magnetization vector can be regarded as a superposition of two
components (Fig. \ref{konfprost}): static (parallel to the
$z$-axis) and dynamic (lying in the $(x,y)$-plane):
\begin{eqnarray}
\vec{M}_{\vec{R}}=M_{S} \hat{k} + \vec{m}_{\vec{R}};   \label{Ms}
\end{eqnarray}
$M_{S}$  is the static magnetization, assumed to be homogeneous
throughout the sample, and vector $\vec{m}$  denotes the dynamic
magnetization, perpendicular to $\vec{M_{S}}$. Similarly, the
dipole field, $\vec{h}_{n'}$, can be resolved into two components:
static, $\vec{h}_{n'}^{s}$  (parallel to the $z$-axis), and
dynamic, $\vec{h}_{n'}^{d}$ (lying in the $(x,y)$-plane):
\begin{eqnarray}
\vec{h}_{n'} =\vec{h}_{n'}^{s}+\vec{h}_{n'}^{d}. \label{statdyn}
\end{eqnarray}
These two components of the dipole field can be easily found from (\ref{eq7a}). By replacing the
third component of the magnetization vector with the static magnetization ({\em i.e.} by
putting $M_{n}^{z} \equiv M_{S}$), and the two other components, $M_{n}^{x}$ and $M_{n}^{y}$,
 with the respective components of the dynamic magnetization, $m_{n}^{x}$  and $m_{n}^{y}$,
 the following formulae are obtained:
\begin{eqnarray}
\vec{h}_{n'}^{s} & = &-\left[ \frac{1}{2 \pi} \sum_{n} D_{n,n'}\right] M_{S} \hat{k}, \label{hstat1}\\
\vec{h}_{n'}^{d}& = & \frac{1}{4 \pi} \sum_{n} D_{n,n'}
\vec{m}_{n}, \label{hdyn1}
\end{eqnarray}
element $D_{n,n'}$ being defined by (\ref{L}).

The magnetic moment dynamics is described by the phenomenological Landau-Lifshitz equation (LL):
\begin{eqnarray}
\frac{\partial \vec{M}_{\vec{R}}}{\partial t} =\gamma \mu_{0}
\vec{M}_{\vec{R}} \times \vec{H}_{eff,\vec{R}}, \label{LL1}
\end{eqnarray}
$\vec{H}_{eff,\vec{R}}$  denoting the effective magnetic field
acting on the magnetic moment in site $\vec{R}$. This effective
field is a superposition of two terms only: the applied field,
$\vec{H}_{0}$, and the field $\vec{h}_{n'}$
 produced by the magnetic dipole system:
\begin{eqnarray}
\vec{H}_{eff,\vec{R}} \equiv \vec{H}_{eff,n'}=\vec{H}_{0} +
\vec{h}_{n'}.
\end{eqnarray}
Considering (\ref{statdyn}), we can write further:
\begin{eqnarray}
\vec{H}_{eff,n'}=\left( H_{0}+h_{n'}^{s}\right) \hat{k}+
\vec{h}_{n'}^{d};
\end{eqnarray}
the above-introduced dipole field components (static and dynamic) being defined by
(\ref{hstat1}) and (\ref{hdyn1}).

The LL equation becomes:
\begin{eqnarray}
\frac{\partial \vec{m}_{n'}}{\partial t} =\gamma \mu_{0}
\left(M_{S}\hat{k}+\vec{m}_{n'} \right) \times \left(\left(
H_{0}+h_{n'}^{s}\right) \hat{k}+ \vec{h}_{n'}^{d}\right).
 \label{LL2}
\end{eqnarray}
We shall solve it using the linear approximation, {\em i.e.}
neglecting all the terms with $\vec{m}$ squared. Assuming the
standard harmonic time-dependence of the solutions: $\vec{m}_{n'}
\sim e^{-i\omega t}$, (\ref{LL2}) becomes:
\begin{eqnarray}
-i\omega \vec{m}_{n'} =\gamma \mu_{0}\left[ M_{S}\hat{k} \times
\vec{h}_{n'}^{d} + \vec{m}_{n'}  \times \left(
H_{0}+h_{n'}^{s}\right) \hat{k} \right]
 \label{LL3}
\end{eqnarray}
or, using the properties of vector product:
\begin{eqnarray}
-i\omega \vec{m}_{n'} =\gamma \mu_{0} \hat{k} \times\left[ M_{S}
\vec{h}_{n'}^{d} - \vec{m}_{n'}  \left( H_{0}+h_{n'}^{s}\right)
\right].
 \label{LL4}
\end{eqnarray}
Through replacing the dipole field dynamic and static components with their explicit expressions
(\ref{hstat1}) and (\ref{hdyn1}) we obtain:
\begin{eqnarray}
-i\omega \vec{m}_{n'} =\gamma \mu_{0} \hat{k} \times\left[ M_{S}
\frac{1}{4 \pi} \sum_{n} D_{n,n'} \vec{m}_{n} -\vec{m}_{n'}
\left( H_{0}-\frac{M_{S}}{2 \pi} \sum_{n} D_{n,n'} \right)
\right],
 \label{LL5}
\end{eqnarray}
or, after bilateral multiplication by $-4\pi(\gamma \mu_{0} M_{S})^{-1}$:
\begin{eqnarray}
i \Omega \vec{m}_{n'} = \hat{k} \times\left[ \vec{m}_{n'}  \left(
\Omega_{H} -2 \sum_{n} D_{n,n'}\right) -\sum_{n} D_{n,n'}
\vec{m}_{n} \right],
 \label{LL6}
\end{eqnarray}
$\Omega$  and $\Omega_{H}$ denoting the {\em reduced frequency} and the {\em reduced field},
respectively, defined as follows:
\begin{eqnarray}
\Omega \equiv \frac{4 \pi \omega}{\gamma \mu_{0} M_{S}}
\mbox{\hspace{0.5cm} and \hspace{0.5cm}} \Omega_{H} \equiv \frac{4
\pi H_{0}}{M_{S}}. \label{omega}
\end{eqnarray}
Two complex variables are now introduced for convenience:
\begin{eqnarray}
m^{\pm}_{n} = m_{n}^{x} \pm i m_{n}^{y}; \label{mpm1}
\end{eqnarray}
with these new variables, (\ref{LL6})  splits into two independent
{\em identical} scalar equations for $m^{+}_{n}$ and $m^{-}_{n}$;
this means we are dealing with magnetostatic waves polarized {\em
circularly}. Therefore, it is enough to consider only one of these
two equations, {\em e.g.}  that for $m^{+}_{n}$:
\begin{eqnarray}
\Omega m^{+}_{n'} = m^{+}_{n'} \left( \Omega_{H} -2 \sum_{n}
D_{n,n'}\right) -\sum_{n} D_{n,n'} m^{+}_{n}.
 \label{LL7}
\end{eqnarray}
The above equation can be rewritten as follows:
\begin{eqnarray}
\Omega m^{+}_{n'} = m^{+}_{n'} \left( \Omega_{H} -2 \sum_{n}
D_{n,n'}-D_{n',n'}\right) -\sum_{n \neq n'} D_{n,n'} m^{+}_{n}.
 \label{LL7b}
\end{eqnarray}
Matrix elements $D_{n,n'}$ mean elements of the dipolar matrix
discussed in Section \ref{theory}; with notations introduced
there, the equation (\ref{LL7b})  becomes:
\begin{eqnarray}
\Omega m^{+}_{n'} = m^{+}_{n'}  \Omega_{n'} -\sum_{\delta=1,2...}
D_{\delta} m^{+}_{n' \pm \delta},
 \label{LL7c}
\end{eqnarray}
where we introduced the following abbreviation denoting {\em the
local field}:
\begin{eqnarray}
\Omega_{n'} \equiv \Omega_{H} -D_{0}- 2 \sum_{\delta=0}{}^{n'}
D_{\delta}.\label{omegan}
\end{eqnarray}
It is also convenient  to introduce at this stage {\em the reduced
static demagnetizing field} - on the analogy of the reduced
external field (\ref{omega}) - by using the definition
(\ref{hstat1}):
\begin{eqnarray}
\Omega_{n'}^{s} \equiv \frac{4\pi}{M_{S}} h_{n'}^{s} =
-2\sum_{n}D_{n,n'}\equiv -2\sum_{\delta=0}{}^{n'} D_{\delta}.
\end{eqnarray}
This allows us to rewrite Eq. (\ref{omegan}) in the form:
\begin{equation}
\Omega_{n'}=\Omega_{H}+\Omega^{d}+\Omega_{n'}^{s},
\end{equation}
with $\Omega^{d} \equiv - D_{0}$ and $\Omega_{n'}^{s}$ meaning the
({\em reduced}) contributions to the local field coming,
respectively, from the dynamical and statical parts of the
demagnetizing field. Note that, according to the assumption made
at the beginning of  this paragraph, the eigenvalues $\Omega$
(being reduced frequencies) correspond to magnetostatic waves
propagating in the direction of the applied field, {\em i.e. along
the central axis} shown in Fig. \ref{plaszczyzny}.

In the remaining part of our work we will be considering only the
cubic-shaped samples, {\em i.e.} starting from this point we
always assume $2L\equiv N-1$. We also assume particular values for
$\mu_{0} H_{0}=0.2T$ and $M_{S}=0.139 \cdot 10^{6} Am^{-1}$ (YIG
magnetization) with resulting value for the reduced field
$\Omega_{H}=14.374$. However, we have to emphasis that selection
of this particular value for $\Omega_{H}$ is not essential for
results to be presented in subsequent sections of this work, since
the distribution of eigenvalues $\Omega$ and profiles of modes
associated with them are not sensitive to the choice of particular
$\Omega_{H}$ value: the particular  value of $\Omega_{H}$ only
sets the {\em whole} spectrum in a given frequency region and if
$\Omega_{H}$ changes the whole spectrum is shifted to another
region, but the {\em relative } distribution of mode
eigenfrequencies remains unchanged.

\section{Mode frequencies and amplitude profiles along the cube central
axis}

We shall investigate magnetostatic excitations in a cube of size
$40a$ ($a$ being the lattice constant). Fig. \ref{plaszczyzny}
shows this cube in a reference system rotated by 90\r{} with
respect to that indicated in Fig. \ref{konfprost} (the $z$-axis,
along which the atomic planes are counted, is now horizontal). The
cube consists of 41 planes normal to the $z$-axis and numbered
with index $n$, ranging from $n$=0 (the left face) to $n$=40 (the
right face). The effective dipole field calculation procedure
applied in the previous paragraphs allows to find the field in
$z$-axis points only, $i.e.$ along the cube \textit{central axis},
passing through opposite cube face centers; this is the idea of
the approximation used throughout this study, and henceforth
referred to as \textit{central-axis approximation} (\textit{CAA}).

The whole sample is assumed to be magnetized uniformly (the
corresponding magnetization value being $M_{S})$ and in the
direction of the external field, applied along the z-axis. The
$z$-axis direction shall be also the only one allowed for
propagation of the magnetostatic waves studied in this paper. With
these assumptions, the problem of motion -- to be solved on the
basis of (\ref{LL1}) -- reduces to a single dimension in the space
of variable $n$; the domain of the investigated motion is the
interval $n\in (0, N-1)$, between two opposite cube face centers
(Fig. \ref{plaszczyzny} shows this interval in close-up).

Fig. \ref{3} presents the discrete spectra of numerically
calculated magnetostatic mode frequencies in a cube of variable
size; the spectrum evolution with increasing $N$, or cube size, is
visualized by the depicted frequency branches, each corresponding
to one mode of a fixed number $m$. The plot shows clearly that
only in the lowest $N$ value range ($N<$50) the frequency spectrum
changes in a significant way; above this range, the frequency
values stabilize at levels independent of $N$. A striking feature
is that the frequencies of the two highest modes, $m=N$--1 and
$m=N$, as well as those of the two lowest ones, $m$ = 1 and $m$ =
2, are pronouncedly separated from the rather uniform ,,band''
formed by the other mode frequencies. In Fig. \ref{4}, showing all
mode profiles in a $40a \times 40a \times 40a$ cube, these
``detached'' modes reveal quite distinct amplitude distributions,
differing from those of the other modes: modes $m=N$ -- 1 and
$m=N$ appear to be of the \textit{bulk-extended }(\textit{BE})
type (with antisymmetrical and symmetrical amplitude distribution,
respectively), whereas modes $m$ = 1 and $m$ = 2 are
\textit{surface-localized} (\textit{SL}), the lower one being
\textit{antisymmetrical}, and the higher one \textit{symmetrical}.
Besides, note that the amplitude arrangement is
\textit{ferromagnetic} in the BE modes, and
\textit{antiferromagnetic} in the SL modes. All the other modes,
within the ,,band'', can be qualified as \textit{localized }($L)$,
their maximum amplitudes localizing in some specific regions
inside the sample; as these localization regions are found to vary
with the mode number, the mode localization appears to depend on
the mode frequency.

To examine this relation in detail, let's note that the band modes
can be divided into two groups: modes $m$ = 3 $\div $ 24 and $m$ =
25 $\div $ 39, showing different localization regions. A
characteristic feature of the first-group modes is a zone of
\textit{zeroing} amplitudes around the sample center; \textit{two}
non-zero amplitude regions are present at both sides of the
central ``dead'' zone. Each of these non-zero regions can be
divided further into two sub-regions: an outer one, situated at
the border, with ferromagnetic arrangement of the amplitudes which
fade towards the sample borders, and an inner sub-region, with
mostly antiferromagnetic arrangement of the amplitudes, reaching a
maximum at a certain point. It is apparent from Fig. \ref{4}b that
the ``ferromagnetic tail'' extends as the mode energy increases,
shifting the neighbouring localization region inside the sample;
finally the localization reaches the sample center, the modes
entering the other group ({\em i.e.} $m$ = 25 $\div $ 39). Typical
for this group, the central localization region tightens around
the sample center as mode energy increases. With respect to the
above-discussed localization properties in both groups, the
second-group modes can be qualified as \textit{center-localized
}(\textit{CL}), and the first-group ones as \textit{non-central
bulk-localized} (\textit{NCBL}) (or, alternatively,
\textit{empty-center bulk localized}).

\section{Two types of mode localization: surface-localized and bulk-localized
modes}

Let's study in detail the NCBL mode properties. An extremely
interesting feature of this group is both amplitude and energy
double degeneration, appearing simultaneously. This effect is
illustrated in Fig. \ref{5}, showing the mode profiles divided
into two groups: symmetrical modes (S) are separated from
antisymmetrical ones (AS) (for simplicity reasons, only $\vert
m^{+}\vert $ profiles are depicted). Note that the profiles of
consecutive mode pairs: $m$=1 (AS) and $m$=2 (S), $m$=3 (AS) and
$m$=4 (S), and so on are \textit{identical}. This amplitude
degeneration in AS-S mode pairs takes place throughout the whole
NCBL group, to pair $m$=23;24 inclusive. Its disappearance above
this pair is easy to explain: as by definition, in antisymmetrical
modes amplitude vanishes in the sample center, a necessary
condition for their symmetrical counterparts to have identical
amplitude distribution is a similar amplitude zeroing in the
center; but this occurs in NCBL modes only. The amplitude
degeneration is associated with that of energy levels; a double
energy level degeneration in the NCBL group is apparent in Fig.
\ref{6}. (When studying Fig. \ref{6}, note once more that the two
highest modes, $m=N$-1 and especially $m=N$, are apparently
separated from the other mode energy levels.)

Getting back to Fig. \ref{5}, note that all the modes have two
ferromagnetic tails et their borders; the extent of these tails
depends on mode frequency. In both the symmetrical and the
antisymmetrical mode groups, the highest-energy mode consists of
two \textit{complete} ferromagnetic tails, each extending from a
sample border to the sample center. In the highest-energy
symmetrical mode, $m=N$, the tails meet in the center with their
maximum amplitudes; in the counterpart antisymmetrical mode,
$m=N$-1, the tails meet in the center with their zero amplitudes.
An interesting thing to note is that each of these two highest
modes can be regarded as the ,,source'' of a mode family formed by
all the other modes within the given symmetry group. The
generation of the lower modes from the initial BE modes, $m=N$ or
$m=N$-1, is based on ferromagnetic tail shrinking: receding from
the sample center, the tails give way to the formation of a new
mode segment, with amplitudes localized in the center (CL modes).
The central localization zone broadens as energy decreases, to
split into two separate (right and left) localization sub-zones,
which migrate towards the borders following the continuously
shrinking ferromagnetic tails. No more non-zero amplitudes appear
in the vacated center, which therefore remains ``dead''; this
corresponds to the formation of NCBL modes. When the non-zero
amplitudes reach the borders (which occurs in the lowest-frequency
modes), the entire inside is dead, and the corresponding modes are
of surface-localized nature.

\section{Size-induced localization effect}

Let's increase the cube size to examine its effect on the mode
energy spectrum and profiles. Fig. \ref{6}b shows a discrete mode
frequency distribution calculated for a cube of size 2$L$=100;
this distribution can be compared to that obtained for a cube of
size 2$L$=40, shown in Fig. \ref{6}a. Besides the obvious increase
in the density of states (resulting from the sample enlargement),
the larger cube retains the basic characteristics observed in the
smaller cube, namely: the double degeneration in the
lower-spectrum modes, its absence in the upper modes, and the
outstanding of the two highest modes. Note that also the mode
frequency range remains virtually unchanged, and that with the
increased density of states, the linear character of the mode
frequency \textit{vs}. mode number dependence is more apparent.

However, this ``invariance'' of the energy spectrum does not
translate into the amplitude profiles. Fig. \ref{7} presents a
juxtaposition of profiles of the five lowest modes and the six
highest ones in cubes of two different sizes: 2$L$=40 and
2$L$=100. The profile invariance is found to be limited to the BE
modes only ($m$=1;2) whose profiles remain unchanged in spite of
the size increase. The other states reveal apparently changed
localization: the NCBL modes in the 2$L$=100 cube, showing
substantially larger dead regions (with respect to the 2$L$=40
cube), are practically of \textit{sub-surface localized} (SSL)
nature; in the CL modes, the localization zone has tightened
around the strict sample center. Thus, we can anticipate that
further size increase shall result in a deepening of the
above-mentioned localization changes, and that consequently, the
magnetostatic mode spectrum in large cubic samples shall consist
of three clearly different mode groups: (a) two highest-frequency
\textit{bulk-extended modes} (\textit{AS} and $S)$, (b)
\textit{center-localized modes} with intermediate frequencies, and
(c) lowest-frequency\textit{ surface-localized modes}. Another way
of illustration of this \textit{size-induced localization effect
}is given on the Fig. \ref{8}.

\section{Internal dipolar field nonhomogeneity as the source of mode
localization}

Obviously, we would like to understand why almost all of the modes
forming the magnetostatic spectrum (except for the two highest
ones) are of localized nature. We can suppose that this property
is due to the long-range character of dipole interactions, as the
latter as known to produce spatially nonhomogeneous
demagnetization field in bounded non-ellipsoid systems \cite{15}.
In fact, this is the case of the cube considered in this paper:
plotted with bold line in Fig. \ref{9}, the local field $\Omega
_{n}$, given by Eq.(\ref{omegan}), shows nonhomogeneous spatial
distribution (along the central axis), reaching a maximum in the
sample center, and fading symmetrically towards the borders. The
magnetostatic mode profiles, plotted in the same energy scale,
have been superimposed on this local field plot. Strikingly, the
shape of the suppressed amplitude area in the bulk resembles
exactly that of the nonhomogeneous distribution of $\Omega _{n}$,
and all the NCBL mode frequencies lie within the local field value
range. The CL mode frequencies coincide with the top of the
$\Omega _{n}$ curve, $i.e.$ the region where the local field
gradient is highest. The BE mode frequencies lie \textit{above
}the $\Omega _{n}$ value range, and one can think that this is why
they are hardly sensitive to the local field variations.

\section{Final remarks and outlooks}

In their recently performed systematic studies of FMR in patterned
submicron rectangular permalloy films, Zhai \textit{et al.}
\cite{16} observed multipeak spectra interpreted as spin-wave
resonance due to the non-uniform magnetization of the studied
samples. We expect a similar study can be performed on cubic
structures as well, with the aim of revealing different types of
the above-described localized modes.

One of the practical aspects of our study concerns the
size-independence of magnetostatic mode frequencies found in
cubic-shaped samples. This is in contrast with the recent report
\cite{17}, where FMR spectra were measured for a series of single
crystal yttrium-iron-garnet (YIG) \textit{films} of thickness
ranging from 35 $\mu $m to 220 $\mu $m. The spectra were
interpreted numerically on the basis of a model similar to ours,
$i.e$. taking into account the strongly inhomogeneous internal
field, but the resonance fields (measured for both in-plane and
perpendicular configurations) were found to be sensitive to the
\textit{film }thickness. This effect is, in fact, not predicted by
the results of our study, performed on \textit{cubic-shaped}
samples: almost identical local field $\Omega _{n}$ profiles
(depicted with bold lines in Fig. \ref{6}) were found inside two
cubes of different sizes, 2$L$=40 and 2$L$=100, and moreover, the
mode frequency spectrum distribution with respect to $\Omega _{n}$
is almost identical in both cases. Therefore, we may conclude that
the above-discussed size-independence of our magnetostatic mode
spectrum resulted from the assumed \textit{cubic} shape of the
studied sample.

A problem to be solved next is the mode localization behaviour
when the sample is ,,deformed'', losing its cubic symmetry. In our
next paper we are going to calculate the magnetostatic mode
spectrum in a rectangular finite thin platelet. Our preliminary
results indicate that the magnetostatic mode localization effects
described here for cubes are present in rectangular platelets as
well. We believe, in fact, that this effect is just underlying the
so-called `bimodal' statistical distribution of the internal
field, recently revealed in finite YIG thin films by
Pardavi-Horvath and Yan \cite{2}, and found to be clustered around
two major components -- a low-field peak corresponding to the
central (volume) part of the sample, and a higher internal field
related to the surface -- which would correspond exactly to CL and
SL modes disscussed in this work. Closer elucidation of this
analogy will be the subject of our subsequent paper.

Another question requiring future investigation is the mode
localization dependence on mode propagation direction. In another
study we are going to examine magnetostatic modes propagating
\textit{perpendicularly} to the applied field.

\begin{acknowledgements}
This work was supported by the Polish Committee for Scientific
Research through the projects KBN - 2P03B 120 23
 and PBZ-KBN-044/P03-2001.
\end{acknowledgements}

\newpage
\begin{figure}
\begin{center}
\caption{ (a) The prism sample considered here; the case in which
the applied field, $\vec{H}_{0}$, sets the magnetic moments in the
direction perpendicular to the prism base ({\em i.e.} along the
$z$-axis). The prism 'thickness' is $(N-1)a$, and the square base
side width is $2La$ ($a$ denoting the lattice constant). The
magnetostatic waves are assumed to propagate along the $z$-axis
({\em i.e.} in the direction of the applied field). (b) The planar
prism model used in our calculations.} \label{konfprost}
\end{center}
\end{figure}

\begin{figure}
\begin{center}
\caption{We consider a cube of edge size 40$a$, magnetized
uniformly along the $z$-axis by the applied magnetic field
($M_{S}$ being the magnetization). Magnetostatic modes are assumed
to propagate only in the direction of the applied field. The
dipolar field entering the equation of motion is calculated
numerically along the central axis (brought into close-up in the
graph).}\label{plaszczyzny}
\end{center}
\end{figure}

\begin{figure}
\begin{center}
\caption{The quantized magnetostatic mode frequencies \textit{vs.}
the cube size, $N$; $m$ indicates the mode number.}\label{3}
\end{center}
\end{figure}

\begin{figure}
\begin{center}
\caption{Numerically calculated magnetostatic mode profiles in a
cubic sample of edge size 40$a$; the profiles are depicted along
the central axis (indicated in Fig. \ref{plaszczyzny}) showing
separately: (a) the dynamical magnetization $m^{+}$ relative
values, and (b) the corresponding absolute values, $\vert
m^{+}\vert $.}\label{4}
\end{center}
\end{figure}

\begin{figure}
\begin{center}
\caption{The modes shown in Fig. \ref{4} are divided here into
symmetrical (right column) and antisymmetrical (left column)
groups. The modes within each group can be regarded as a family
generated by the \textit{bulk-extended} top mode: with lowering
mode frequency, the ferromagnetic mode tails shrink, resulting in
the appearance of three mode localization types:
\textit{central-bulk}, \textit{non-central} and \textit{surface}
localization.}\label{5}
\end{center}
\end{figure}

\begin{figure}
\begin{center}
\caption{The quantized magnetostatic mode frequencies (square
points) \textit{vs.} the mode number, $m$; note the double
degeneration of the lower-spectrum modes. Note also that the
frequencies of modes $m=N$ and $m=N-1$ are clearly separated from
those of the remaining modes. The bold line represents the spatial
distribution of the local dipolar field $\Omega _{n}$ values along
the central axis. The cube edge size is (a) 2$L=N-1=$40, and (b)
2$L=N-1=$100.}\label{6}
\end{center}
\end{figure}

\begin{figure}
\begin{center}
\caption{Juxtaposition of magnetostatic mode profiles in two cubes
of different sizes. Note that when the cube size increases, the
lower modes tend to localize in the surface region, while the
upper modes become strongly localized around the cube center; the
two highest modes remain practically unchanged.}\label{7}
\end{center}
\end{figure}

\begin{figure}
\begin{center}
\caption{An illustration of the size-induced localization effect.
When the cube size increases, the localization regions of the
localized modes ($m=N$-6, 5 and 1) narrow down
substantially.}\label{8}
\end{center}
\end{figure}

\begin{figure}
\begin{center}
\caption{Magnetostatic mode spectrum in a cube of size 40$a$
confronted with the spatial distribution of the local dipolar
field $\Omega_{n}$ (bold line) along the cube central axis. A
striking feature is that the local field line borders the
``empty'' region where mode amplitudes are suppressed. The
bulk-extended mode frequencies lie \textit{above} the local field
range.}\label{9}
\end{center}
\end{figure}


\begin{thebibliography}{50}
\bibitem[*]{correspond} Corresponding author: H. Puszkarski, Surface
Physics Division, Institute of Physics, Adam Mickiewicz
University, ul. Umultowska 85, 61-614 Poznan, Poland;
\textit{Email address}: henpusz@amu.edu.pl
\bibitem{1}D. Shi, B. Aktas, L. Pust, F. Mikailov (Eds.),
\textit{Nanostructural Magnetic Materials and Their Applications},
Lectures notes in physics: Vol. 593; (Springer-Verlag, Berlin
2002).
\bibitem{2}M. Pardavi-Horvath and Jijin Yan, IEEE Trans. Magn,
\textbf{39} (2003) 3154.
\bibitem{3}Toshinobu Yukawa and Kenji Abe, J. Appl. Phys. \textbf{45} (1974) 3146.
\bibitem{4}Z. Zhang {\em et al.}, Appl. Phys. Lett. \textbf{68} (1996) 2005.
\bibitem{5}G.P. Berman, F. Borgonovi, Hsi-Sheng Goan, S.A. Gurvitz, and
V.I. Tsifrinovich, Phys. Rev. \textbf{B67} (2003) 094425.
\bibitem{6}J. Jorzick, S.O. Demokritov, B. Hillebrands, M. Bailleul, C.
Fermon, K.Y. Guslienko, A.N. Slavin, O.V. Berkov and N.L. Gorn,
Phys. Rev. Lett. \textbf{88} (2002) 47204.
\bibitem{7}J.P. Park, P. Eames, D.M. Engebretson, J. Berezovsky and P.A.
Crowell, Phys. Rev. Lett. \textbf{89} (2002) 277201.
\bibitem{8}D.V. Berkov, N.L. Gorn and P. G\"{o}rnert, phys. stat. sol.
(a) \textbf{189} (2002) 409.
\bibitem{9}S. Tamaru, J.A. Bain, R.J.M. van de Veerdonk, T.M. Crawford,
M. Covington and M.H. Kryder, J. Appl. Phys\textbf{. 91} (2002)
8034.
\bibitem{10} R. Damon and J. Eshbach, J. Phys. Chem. Solids \textbf{19
}(1961) 308.
\bibitem{11}K.Y. Guslienko, R.W. Chantrell and A.N. Slavin, Phys. Rev.
\textbf{B 68} (2003) 24422.
\bibitem{12}H.J.G. Draaisma and W.J.M. de Jonge, J. Appl. Phys\textbf{.
64} (1988) 3610.
\bibitem{13}E.Y. Vedmedenko, H.P. Oepen, J. Kirchner, J. Magn. Magn.
Matter \textbf{256} (2003) 237.
\bibitem{14}G. Gubbiotti {\em et al.} J. Appl. Phys. \textbf{93} (2003) 7595
and 7607.
\bibitem{landau} L.D. Landau and E.M. Lifshitz, {\em The Classical Theory of
Fields} (Pergamon Press, Oxford, 1975), p. 119.
\bibitem{15}Norberto Majlis, \textit{The Quantum Theory of Magnetism},
World Scientific Publishing Co., Singapore 2000.
\bibitem{16}Y. Zhai, J. Shi, X.Y. Zhang, L. Shi, Y.X. Xu, H.B. Hung,
Z.H. Lu and H.R.Zhai, J. Phys.: Cond. Matter 14 (2002) 7865
\bibitem{17}Martha Pardavi-Horvath, Bela Keszei, Janos Vandlik and
R.D. McMichal, J. Appl. Phys. 87 (2000) 4969
\end{thebibliography}
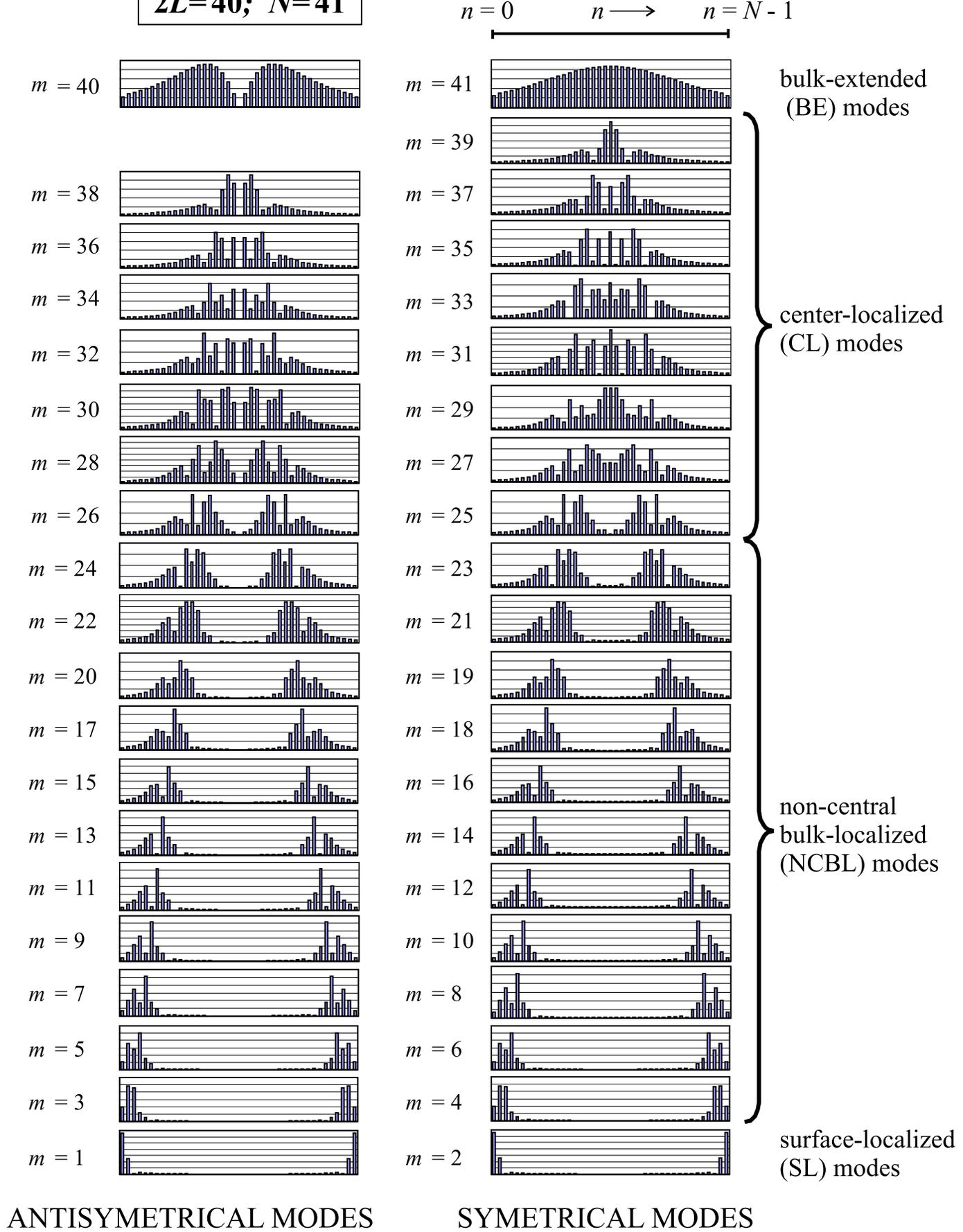
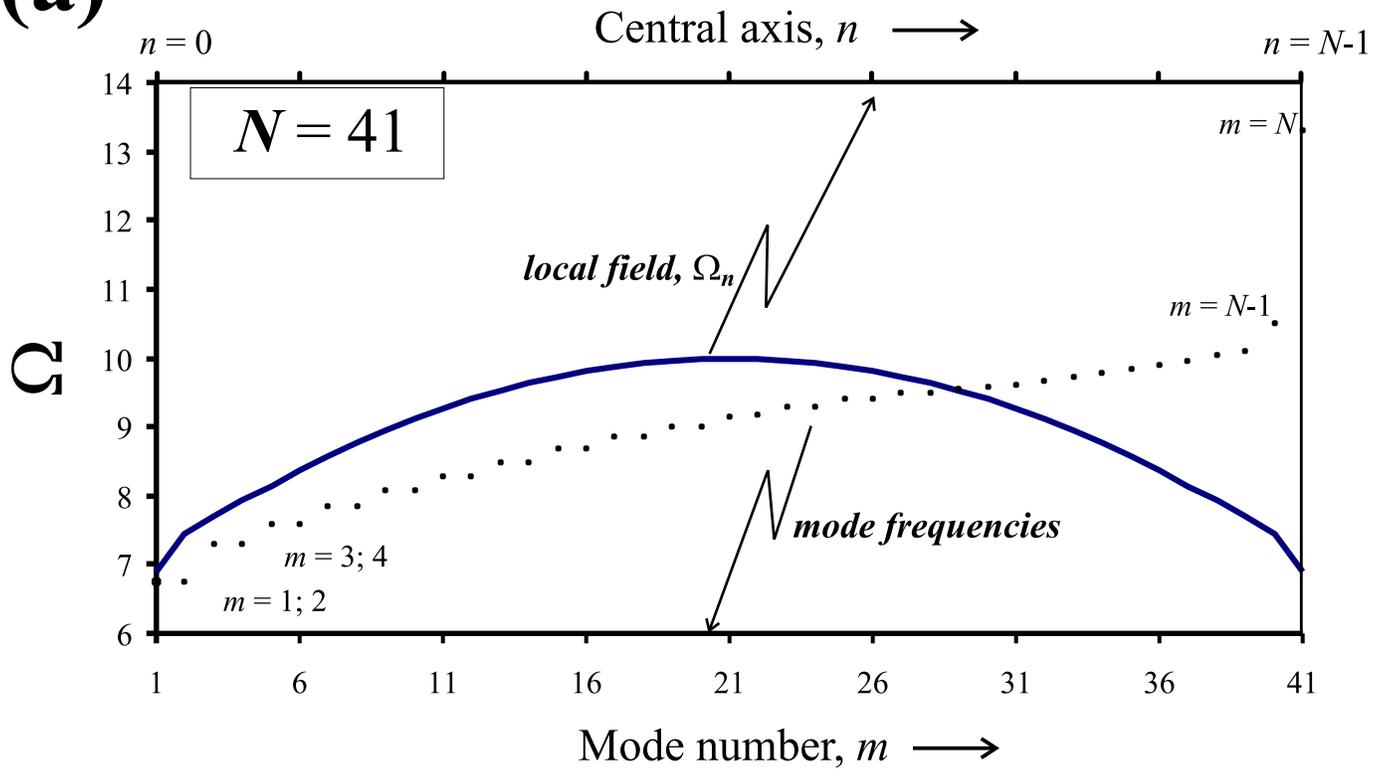
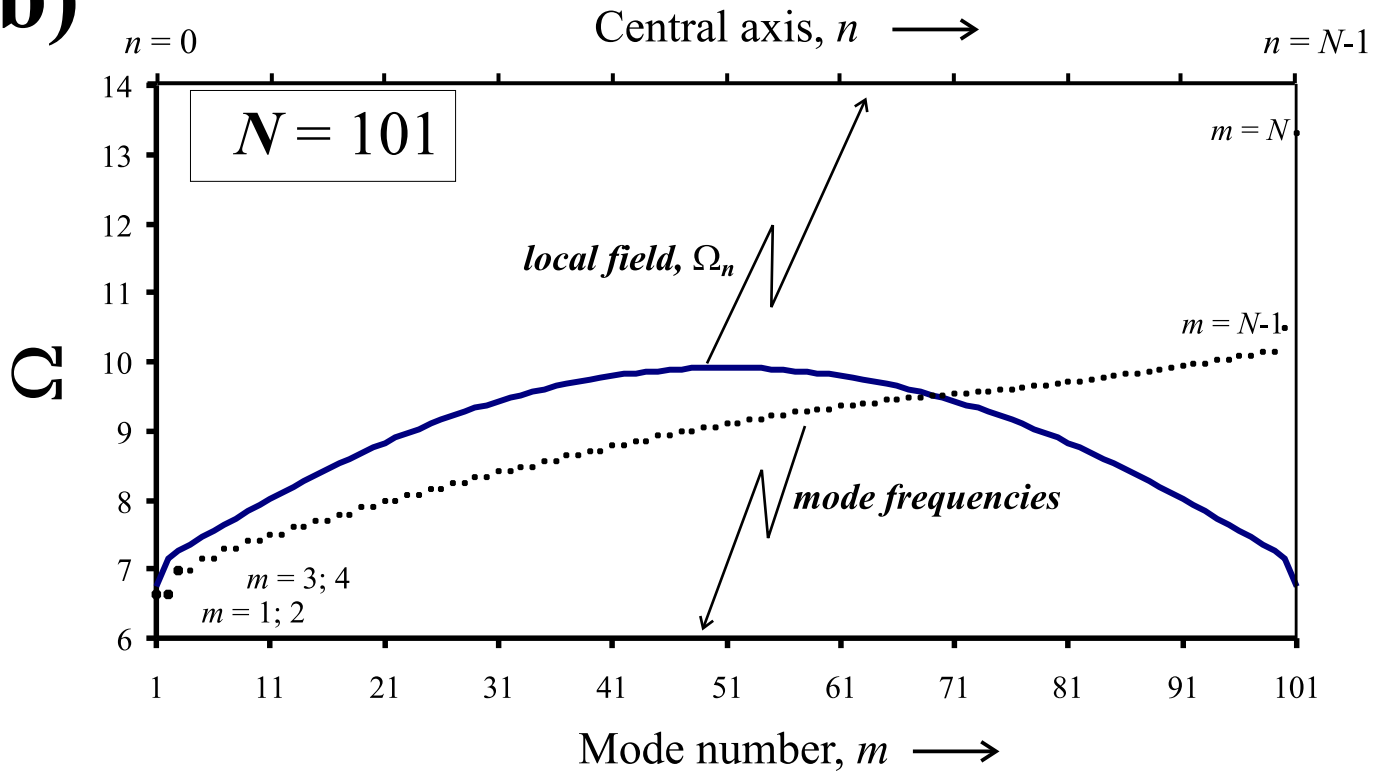
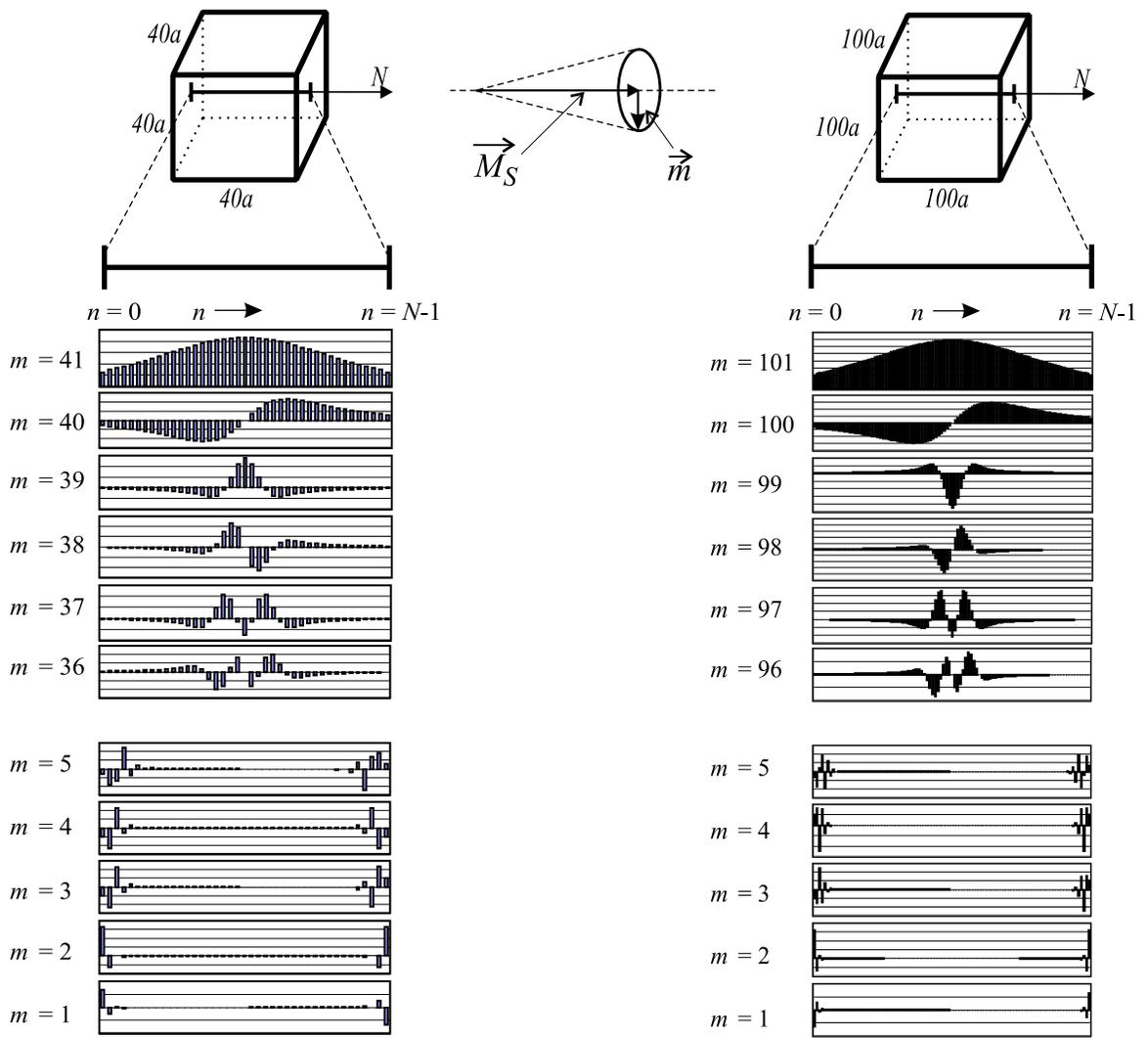
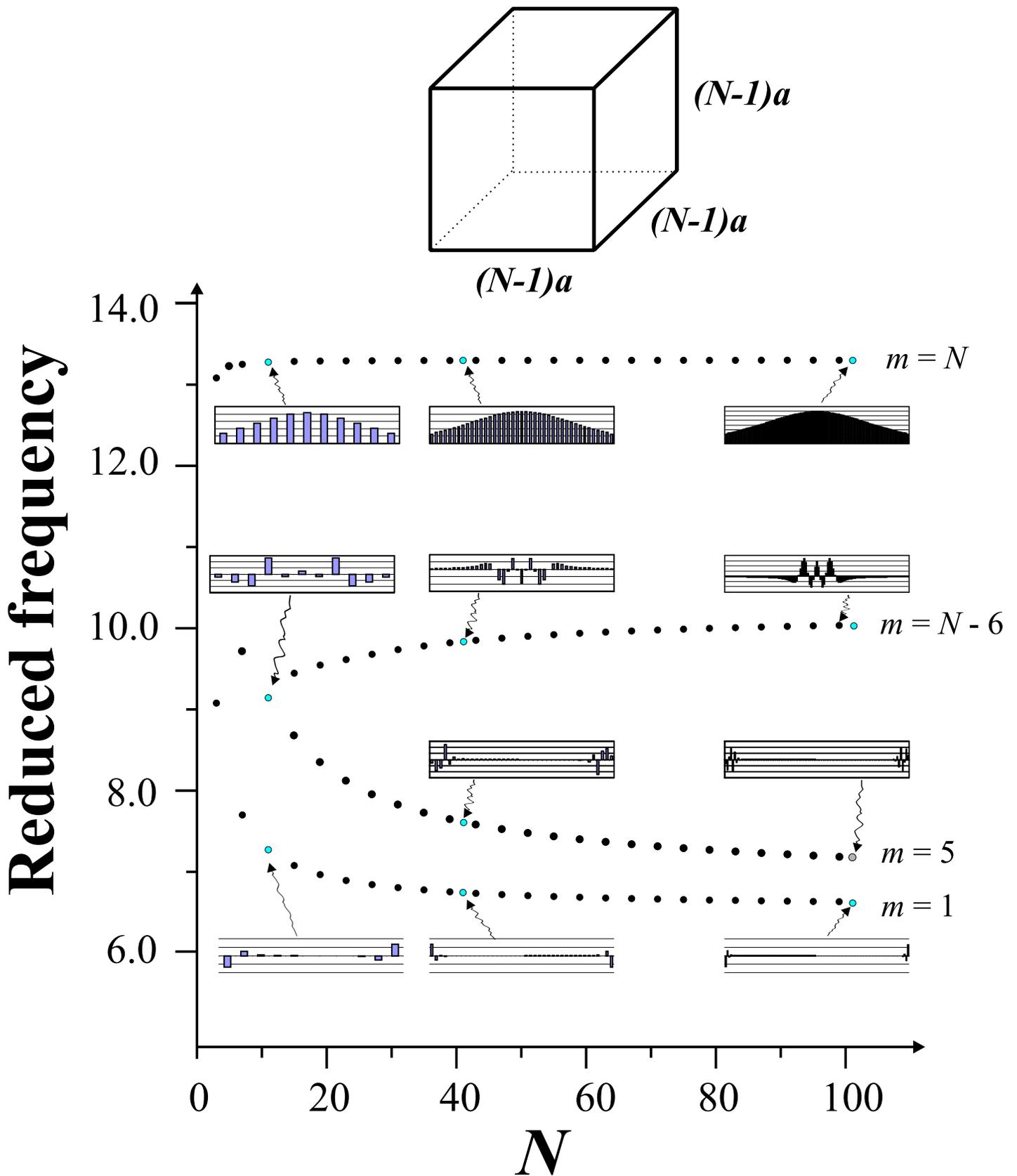
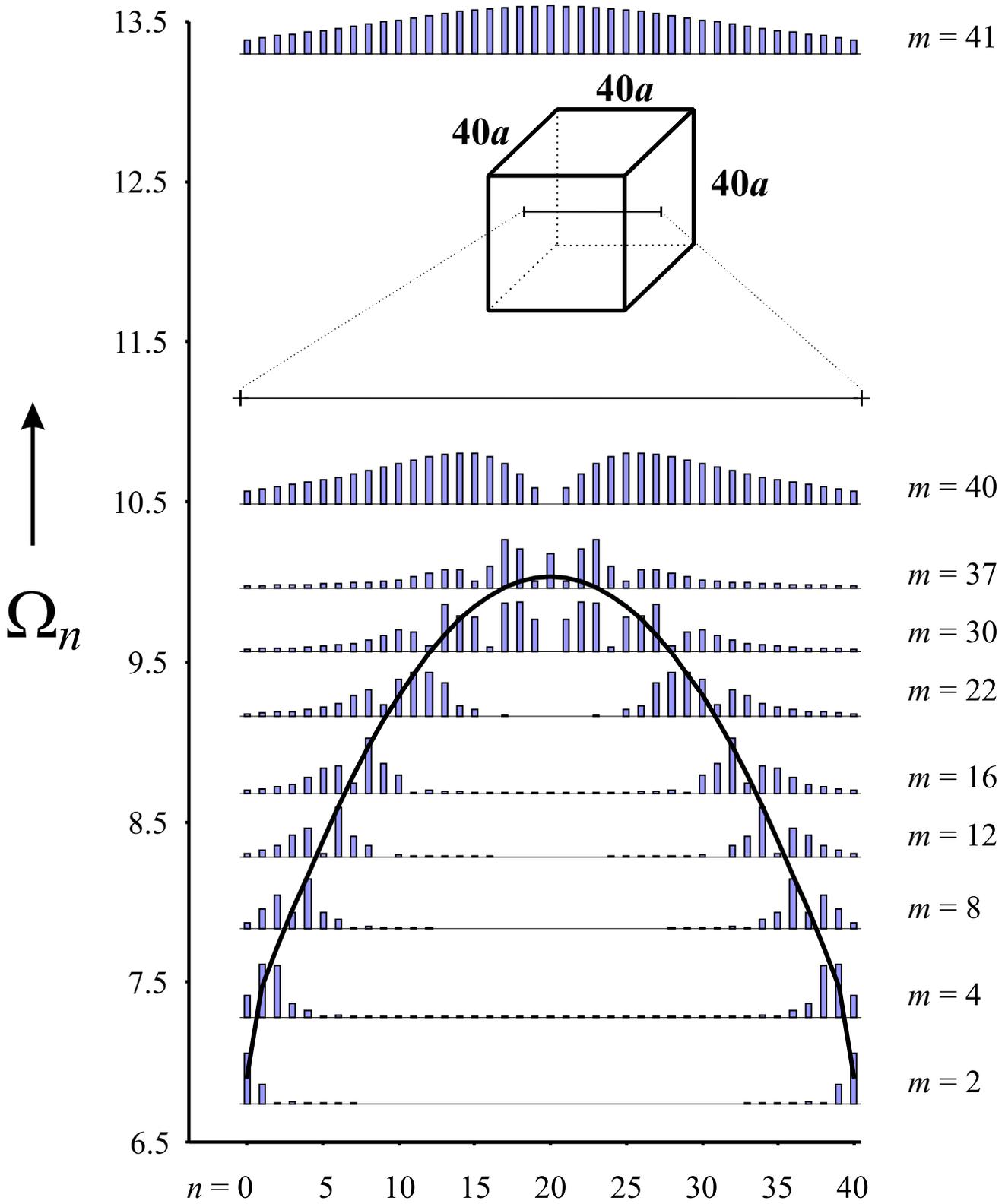
\end{document}